\documentclass[12pt]{article}
\usepackage{epsfig}

\def\beq{\begin{equation}}
\def\eeq#1{\label{#1}\end{equation}}
\def\eeqn{\end{equation}}

\def\beqa{\begin{eqnarray}}
\def\eeqa#1{\label{#1}\end{eqnarray}}
\def\eeqan{\end{eqnarray}}

\let\bar=\overbar

\def\eg{{\it e.g.}}

\def\Dslash{\not{\hbox{\kern-4pt $D$}}}
\def\dslash{\not{\hbox{\kern-2pt $\del$}}}

\def\ee{e^+e^-}

\def\msb{{\bar{\ssstyle M \kern -1pt S}}}

\def\Title#1{\begin{center} {\Large {\bf #1} } \end{center}}

\newcommand{\be}{\begin{equation}}
\renewcommand{\ee}{ \end{equation}}
\newcommand{\nn}{\mbox{} \nonumber \\ \mbox{} }
\newcommand{\ba}{\begin{eqnarray}}
\newcommand{\ea}{\end{eqnarray}}

\renewcommand\eg{\textit{e.g.\ }}

\begin{document}

%\title{ Dynamics of Relativistic  Magnetic Explosions}
\Title{Electromagnetic Outflows and GRBs}
\bigskip\bigskip

%+\addtocontents{toc}{{\it M. Lyutikov}}
%+\label{LyutikovStart}

\begin{center}
{ \bf \large MAXIM LYUTIKOV $^{1,2,3}$, ROGER BLANDFORD $^4$}
\end{center}

%\author{Maxim Lyutikov  $^{1,2,3}$}
%\author{Roger Blandford $^4$}
\begin{raggedright}  

{\it $^1$ Physics Department, McGill University, 3600 rue University
Montreal, QC,  Canada H3A 2T8; \\
$^2$  Massachusetts Institute of Technology,
77 Massachusetts Avenue, Cambridge, MA 02139; \\
 $^3$ CITA National Fellow}\\
{\it $^4$ Theoretical Astrophysics, California Institute of Technology,
Pasadena, California 91125}

\bigskip\bigskip
\end{raggedright}

\begin{abstract}
We  study the dynamics of relativistic electromagnetic
explosions as a possible mechanism for the production of Gamma-Ray Bursts.
We propose that a rotating relativistic stellar-mass
progenitor loses much of its spin energy in the  form of an 
electromagnetically-dominated outflow.
 After the 
flow becomes optically thin, it forms a relativistically expanding,
non-spherically symmetric magnetic  bubble -  a ''cold fireball''.
We analyze the structure and  dynamics
 of  such a  cavity in the force-free approximation. 
During relativistic expansion,
 most of the magnetic 
 energy in the bubble
is concentrated in a thin shell near its surface (contact discontinuity).
We suggest that either the polar current or the shell currents
become unstable  to electromagnetic 
instabilities at a radius $\sim10^{16}$~cm. This leads to
acceleration of  pairs and causes the
$\gamma$-ray emission. At a radius $\sim10^{17}$~cm, the momentum
contained in the electromagnetic shell  will have been largely transferred
to the surrounding blast wave propagating into the circumstellar medium.
Particles accelerated at the fluid shock may combine with electromagnetic 
field from the electromagnetic shell to produce the afterglow emission.
%\keywords{gamma-rays: burster -  magnetic fields }
\end{abstract}
\vskip -.3 truein

\vskip -.3 truein
\section{Introduction}
Most contemporary explanations of Gamma Ray Bursts (GRBs) attribute them 
to high entropy ``fireballs'' which convert their energy 
 into a matter-dominated (baryonic) jet within which 
relativistic electrons and electromagnetic field are subsequently
re-created (e.g. Piran 1999). In our opinion,
 there are  a number of problems with this 
scenario.
 Besides the commonly recognized
problems of  low efficiency and
 baryon mass and magnetic-field  fine-tuning, the creation of the
 plasma-dominated
flows is problematic.
 It is usually envisaged that the energy is released 
electromagnetically (e.g., MacFadyen \& Woosley  1999; Kim et al. 2002; 
Wheeler et al. 2000) and is dissipated right away into 
``lepto-photonic'' plasma. 
 Yet, there is no clear mechanism for 
transforming magnetic energy into a hot  plasma 
on an outflow time-scale. 
 Also, the baryonic jets are inferred to have 
Mach numbers $\sim300$ and to maintain a ratio 
of bulk to internal energy $>10^5$, (or equivalently to maintain a flow
that deviates locally by angles of less than $\sim10'$),
while dissipating heavily through internal shocks. 
  
We explore an alternative model for GRBs in which
the energy remains in electromagnetic form all the way from the
its origin to the sites of $\gamma$-ray and afterglow emission.
%Earlier versions of these ideas are presented in, e.g., Usov (1992), Lyutikov
%(2001), Blandford (2002).
\vskip -.2 truein

\vskip -.2 truein
\subsection{The case for electromagnetic energy transport}
Several arguments point to the importance of electromagnetic field in 
producing GRBs: 
(i) Magnetic fields naturally collimate outflows and have been 
invoked in all other jet outflows.
(ii) Magnetic outflows  are ``clean'' --  
instead of requiring a finely tuned baryon fraction, baryons are 
may be absent completely.
(iii) There is no need to convert Poynting flux into fluid energy flux
and then back again.
(iv)  Electromagnetic energy is ``high quality''  -- it 
is in a low entropy form and can be
efficiently dissipated through current instabilities during particle
acceleration.
(v) Electromagnetic dissipation is intrinsically intermittent 
(c.f. solar flares); the observed GRB variability need not be 
tied to the source and can consequently arise at much larger radii than 
in internal shock models, thereby relaxing the $\gamma$-ray opacity constraint.
(vi) ``Standard candle'' --
 simple electromagnetic models offer an explanation for the 
surprisingly narrow reported burst energy distribution
--  approximately 
the same burst energy will be inferred independent of the viewing
angle (c.f. Frail et al. 2001; 
Panaitescu et al. 2002; Lazzati et al. 2002).
(vii) Ultrarelativistic electromagnetic outflows are formally
subsonic which implies that they are more naturally 
established than hypersonic, fluid jets.
(viii) There are   astrophysical examples of
collimated Poynting  flux dominated outflows 
 (pulsars, AGNs, black hole candidates).
\vskip -.2 truein

\vskip -.2 truein
\subsection{Sources of Electromagnetic Outflow}
We assume that the GRB source (sometimes referred to as
``millisecond magnetar'', Usov 1992) is a strongly magnetized 
neutron star (e.g., Kouveliotou et al., 1998)
or a stellar mass black hole-accretion torus 
 with mass $M\sim1-10M_{\odot}$, size
given by the effective light cylinder radius
$r_0\sim10$~km, angular velocity
$\Omega \sim r_0/c \sim 10^4 $~rad s$^{-1}$ and magnetic field strength
$ B_0 \sim  10^{15}$~G.
Such super-strong magnetic fields may be generated by 
$\alpha-\omega $ dynamos (e.g., Thompson \& Murray, 2001), or just by 
differential rotation (e.g., Klu\'zniak \& Ruderman, 1998).
A rotational energy,
$E\sim 0.1Mr_0^2 \Omega^2 \sim 10^{52}$~erg -- a one parameter quantity,
is available to power GRB bursts (intermediate mass black  holes can 
supply more energy.) The field is chosen to produce 
a characteristic electromagnetic power 
$L\equiv\Delta\Omega L_\Omega\sim{B_0^2r_0^6\Omega^4/c^3}\sim10^{50}$~erg 
s$^{-1}$,
comfortably larger than that inferred for GRBs and giving a source 
lifetime $\sim100$~s matched to long bursts.
\vskip -.2 truein

\vskip -.2 truein
\subsection{The Electrical Circuit}
Independent of the source, we suppose that the combination 
of magnetic flux $\Phi\sim B_0r_0^2\sim10^{27}$~G cm$^2$ and angular velocity
combine to create a unipolar inductor with EMF ${\cal E}\sim0.1\Omega\Phi/c
\sim10^{22}$~V and that the axisymmetric part of the 
electromagnetic field dominates at large radius. An equivalent and useful way
to think about this is to say that there is a strong, quadrupolar current 
distribution outward along the axes and inward along the equator (or 
{\it vice versa}). Under electromagnetic conditions, the effective impedance
is roughly that of free space -- $\sim $ 100~$\Omega$ is fine for estimates -- 
so the current flowing around the circuit is $I\sim10^{20}$~A and the power
that is delivered to the ``load''`is $L\sim {\cal E}I
\sim10^{49}$~erg s$^{-1}$.  (This 
interpretation of the load resistance is valid even if there is no 
dissipation and the electromagnetic Poynting flux just propagates 
away from the source with little reflection.) Our proposal
differs from the conventional interpretation principally through 
the assumption that the current flows all the way out 
to the expanding blast wave, rather 
than completes close to the source (c.f. Fig.~2).
In addition, we make the conjecture that the complex magnetic
field geometry, that must be present within $r_0$, sorts itself out 
and becomes axisymmetric and 
primarily toroidal for $r>>r_0$, by analogy to what is observed
in the quiet solar wind. This can occur completely without (\eg Blandford 2002)
or with partial  (\eg Lyutikov 2002b) dissipation of magnetic field. 
\vskip -.2 truein

\vskip -.2 truein
\subsection{Lepton Loading}
The sources that are envisaged generally 
release energy though baryonic, leptonic and electromagnetic 
channels. The first is usually parameterized by the initial
baryon rest mass fraction of the total energy density
$\eta$, which we shall assume to be quite small. It is also useful to define
a quantity $\sigma$ which is the ratio of the electromagnetic to the 
total matter energy density.  

Somewhat paradoxically, it gets harder to convert electromagnetic energy 
directly to pair plasma the stronger the field becomes. 
The reason is that plasma tries to short out any electric field
along magnetic field on  a very short time scale, a few Langmuir periods.
Even if there is not enough plasma to enforce ${\bf E}\cdot{\bf B}=0$
conditions the required amount of pairs will be created 
through vacuum breakdown.  Hoverer, pair 
 creation is not likely to drain all the potential EMF since  the
 potential differences
required to create pairs through a vacuum breakdown are typically $\sim1$~GV
and never more than $\sim1$~TV, which is  orders of magnitude smaller
than the EMF  required to account for ultra-relativistic
outflows  $\sim10$~ZV in  GRBs.
When the field strength is as high as $\sim10^{15}$~G, 
the minimum density of plasma needed to short out the electric 
field -- the Goldreich-Julian density $n_{{\rm GJ}}$-- is tiny in comparison 
with the equipartition density $n_{{\rm eq}}$.
Put another way,
$\sigma_0<\sigma_{{\rm max}}\sim n_{{\rm eq}}/n_{{\rm GJ}}
\sim{\cal E}/\Delta V_{{\rm vac}}\sim\omega_{{\rm G}}/\Omega
\sim10^{16}$. 

The actual value of $\sigma_0$ is model-dependent. At one extreme,
if there is an evacuated  
neutron star or black hole magnetosphere, only enough pairs need be 
created to supply the 
necessary space charge and current so that $\sigma_0\sim\sigma_{{\rm max}}$. 
At the other, there
might be a weakly magnetized hot torus radiating neutrinos
with energy above $\sim5$~MeV so that $\sigma<<1$ and the standard 
fireball model would apply. All intermediate 
values are possible (e.g., dissipation of large scale magnetic field through
pair creation is likely to
 give $\sigma_0 \sim {\cal{EMF}} /  \Delta V_{break down} \sim 10^{10}$).
 If we suppose that the flow is 
initially electromagnetically dominated, with $\sigma \gg 1$, 
then about the only way that it seems possible to 
create entropy directly and reduce $\sigma$ appreciably is for
an electromagnetic turbulence cascade to develop,
operating down to wavelengths, $\lambda_{{\rm min}}$,
so small that electromagnetic energy can be dissipated in accelerating
pairs, somewhat analogous to the viscous dissipation that terminates 
a fluid turbulence spectrum. If this really can operate then it is hard 
to see how more than a few percent of the electromagnetic energy will be 
dissipated on a source dynamical  time scale, more specifically, $\sigma\sim
\ln(r_0/\lambda_{{\rm min}})\sim100$, and the flow will remain
electromagnetically dominated.

In fact, even if there is an appreciable pair content close to the
source (but still $\sigma \gg 1$), 
this will quickly become irrelevant because the pairs will 
annihilate when their temperature falls to $\sim20$~keV, just like in the 
early universe.
The Thomson optical depth of the plasma will then drop to a low enough 
value that the photons can escape and decouple from the flow.
If this happens in GRBs then a weak $\gamma$-ray precursor is predicted
(Lyutikov \& Usov 2000; Fig.~1) and this may have been observed
(Preece 2002). 
\vskip -.2 truein

\vskip -.2 truein
\subsection{Outflow Phases}
Before discussing a few of the details, we should distinguish the principal
stages through which the outflow passes in our GRB model. 

1. {\bf Flow formation ($r<r_0$)} There must be a quasi-steady source of 
electromagnetic power within
$r_0\sim10$~km. For a long burst, this should last for $\sim10^6$
dynamical times.  The initial temperature of the pair plasma
 is $T_0\sim3(\sigma_0/100)^{-1/4}$~MeV 
and the associated optical depth 
$\tau_{{\rm T}}\sim (B/B_{{\rm L}})(n/n_{{\rm GJ}})$, where
$B_{{\rm L}}\sim5\times10^{15}$~G is the Larmor field
for which the Larmor radius of a mildly relativistic
electron equals the classical electron radius.   
The flow is initially sufficiently optically thick to trap the radiation.

2. {\bf  ``Warm''  acceleration ($r_0<r<r_{{\rm thin}}$)}.
The plasma is accelerated by the electromagnetic field 
and the bulk Lorentz factor increases linearly with
radius until the flow becomes optically thin at $r_{{\rm thin}}\sim 
10^8(\sigma_0/100)^{-1/4}$. A small admixture of baryons will extend
$r_{{\rm thin}}$ but it cannot delay the escape of photons by much.

3. {\bf Electromagnetic bubble}.
 Beyond the photosphere the radiation and $e^\pm$ decouple
 from magnetic
 field - loading drops by 7 orders of magnitude, leaving plasma
strongly magnetized  $\sigma \sim 10^9$. At this moment 
the flow becomes a relativistically expanding 
 magnetically dominated bubble. Several stages of the bubble
expansion can be identified:

3.1.  {\it Magnetic acceleration} ($r_{{\rm thin}}<r<ct_{{\rm s}}$).
%In this phase, the flow can accelerate at a rate that depends upon 
%detailed microphysical effects in the flow.  
The radial
Lorentz factor of the frame in which the electric vector vanishes 
can reach $\Gamma\sim10^4$, as long as it does not exceed $\eta^{-1}$. 
A filled electromagnetic bubble will be produced with polar and axial current
that complete as a ``Chapman-Ferraro'' current along a relativistically 
expanding contact discontinuity (CD) that 
separates the bubble from the swept up circumstellar medium. 
The magnetic field 
will be primarily toroidal and the electric field 
should be primarily poloidal in the frame of the explosion. This phase will 
continue until $\sim c t_{\rm s} 
\sim3\times10^{12}$~cm, when the source switches off. 

3.2. {\it ``Coasting'' electromagnetic shell ($ct_{{\rm s}}<r<r_{{\rm sh}}\equiv
(L_\Omega t_{{\rm s}}^2/\rho c)^{1/4}$)}.
At this time, the bubble becomes 
a relativistically expanding shell of thickness $\sim ct_{{\rm s}}$.  
The shell still contains toroidal magnetic field but the current now 
detaches from the source and completes along the shell's inner surface.
At this stage the  CD  is constantly re-energized by the
fast-magnetosonic waves propagating from the central source 
so that the Lorentz factor of the
 CD is   $\Gamma \sim (L_\Omega/\rho c^3)^{1/4}  r^{-1/2}$ (in a constant
density medium) or  $\Gamma \sim {\rm const}$ (in a $\rho \sim r^{-2}$ wind).
This stage is   limited by one
dynamical time scale  $t_{dyn}
\sim 2 t_{\rm s} \Gamma^2$.

3.3. {\it Self-similar  electromagnetic shell} ($r_{{\rm sh}} <r<r_{{\rm NR}}\equiv(L_\Omega t_{{\rm s}}/\rho c^2)^{1/3}$).
After one dynamical time scale all the regions of the bubble come
into a causal contact -- most of the 
 waves reflected from the CD have propagated throughout the bubble.
As the expanding bubble performs
 a work on the surrounding medium its total energy
decreases;
the amount of energy that remains in the bubble at the self-similar
stage needs to be calculated numerically.
 Most energy in the bubble is still 
concentrated in a thin
shell with $\Delta R  \sim R/ \Gamma^2$ near the surface of the bubble 
which is moving according to $\Gamma \sim \sqrt{E_\Omega / \rho c^2}\, r^{-3/2}$
(in a constant density medium), or $\Gamma \sim r^{-1/2} $
 (in a $\rho \sim r^{-2}$ wind). 
Interestingly, the   structure of the {\it magnetic}
 bubble resembles at this stage the structure of the {\it hydrodynamical}
 relativistic blast wave wave (Blandford\&McKee 1976).
After the  bubble came into a causal contact it starts to evolve in lateral
direction trying to adjust magnetic pressures. This evolution may
be accompanied by electro-magnetic instabilities which lead to  particle
acceleration and 
$\gamma$-rays  production. 
The result of the  lateral energy redistribution
is  a creation of  an anisotropic expansion,
 being faster at a given time
 In addition, 
$\gamma$-rays may be produced throughout the CD surface
 due to development of current
instabilities or inertial acceleration (Smolsky \& Usov 1996).

4. {\bf Relativistic blast wave ($r_{{\rm sh}}
<r<r_{{\rm NR}}\equiv(L_\Omega t_{{\rm s}}/\rho c^2)^{1/3}$)}. 
As the bubble expands its energy is gradually transfered 
 to the preceding forward shock wave. Most efficiently this transfer occurs
at the end of the coasting phase (at the coasting phase $dE/dt \sim r$). 
During the self-similar phase 
 the energy in the magnetic bubble decreases slowly (logarithmically 
in times since at this stage  $dE/dt \sim 1/r$),
 so that  the relativistic blast wave stage is coexistent with
the self-similar stage of the magnetic bubble.
At the forward shock wave
relativistic particles are accelerated producing
the observed  afterglow. Magnetic flux is either incorporated
into the blast wave from the bubble or, perhaps, amplified at the 
relativistic shock front. This is the afterglow phase which 
usually becomes unobservably faint after the expansion speed becomes 
mildly relativistic at $r\sim10^{18}$~cm.

5. {\bf Non-relativistic blast wave ($r_{{\rm NR}}<r$)}.
Eventually the source will expand non-relativistically and become
more spherical with time, resembling a normal supernova remnant.
\vskip -.2 truein

\vskip -.2 truein
\section{Dynamics of magnetic explosions}
\label{Dyna}
\subsection{Relativistic MHD vs force-free formalism}
The conventional method for handling relativistic, magnetized flows
is to use the relativistic extension of regular, non-relativistic
magnetohydrodynamics (RMHD). However, there is a simpler extension 
which is appropriate when the plasma is sufficiently tenuous that its
inertia can be ignored, though sufficiently dense that it can supply the
space charge and current density
($1<<\sigma<<10^{16}$). Under these circumstances, we adopt the relativistic 
force-free (RFF) approximation 
\be
\rho{\bf E}+{\bf j}\times{\bf B}=0
\label{ffa}
\ee
This implies that ${\bf E}\cdot{\bf B}$ and its temporal derivative 
can be set to zero. In addition, 
we restrict our attention to the case $E<B$. Eq.~(\ref{ffa}) allows us to
define an electromagnetic velocity 
${\bf v}={\bf E}\times{\bf B}/B^2$ perpendicular 
to the magnetic field (it is not useful to define a component
along the magnetic field).

Now, Maxwell's equations can be written 
\be
{\partial{\bf E}\over\partial t}=\nabla\times{\bf B}- 4 \pi {\bf j}, \,\,
{\partial{\bf B}\over\partial t}=-\nabla\times{\bf E},
\ee
where we can use $\partial({\bf E}\cdot{\bf B}/\partial t=0$ to derive
\be
{\bf j}=\left[({\bf E}\times{\bf B})\nabla\cdot{\bf E}+
({\bf B}\cdot\nabla\times{\bf B}-{\bf E}\cdot
\nabla\times{\bf E}){\bf B}\right] / B^2.
\label{FF}
\ee

This RFF equation set represents  a simple,
evolutionary dynamical system (Uchida 1997).
When one includes the constraints ${\bf E}\cdot{\bf B}=
\nabla\cdot{\bf B}=0$, there are four, independent 
electromagnetic variables to evolve and four characteristics
along which information is propagated.  In the linear 
approximation, these correspond to forward and backward
propagating fast and intermediate wave modes
with phase speeds $c$ and  $c\hat{\bf k}\cdot\hat{\bf B}$ respectively. 

RFF dynamics can be developed in a manner
that is quite analogous to regular hydrodynamics, with the anisotropic Maxwell
stress tensor taking the place of the regular pressure and the electromagnetic
energy density playing the role of inertia. There is an important difference,
though, in that the existence of a luminal fast mode means that 
electromagnetic ``flows'' do not become truly ``supersonic''. 

The RMHD approach, which must be used in the presence of significant
inertia is based upon the fluid velocity ${\bf u}$. The electric field
is generally supposed to be related to the magnetic field through an infinite
conductivity Ohm's law, ${\bf E}+{\bf u}\times{\bf B}=0$ and there are
seven independent variables to evolve, with seven independent characteristics
(two fast, two intermediate, two slow and one adiabatic) to track.   
\vskip -.2 truein

\vskip -.2 truein
\subsection {Suppression of lateral dynamics}
Relativistic flows do not collimate easily on account of relativistic, 
kinematic effects.
A simple way to see this is by analyzing the MHD force balance equation
in the $\theta$ direction
\be
\partial_t[(w+b^2)\gamma^2 \beta_\theta]+{1\over{r\sin\theta}} 
\partial_{\theta}[\sin\theta(p+b^2/2)]-\cot\theta{{p-b^2}\over r}=0
\label{later}
\ee
where $w$ is the enthalpy and $b=(B^2-E^2)^{1/2}$ 
is the magnetic field in the comoving frame.
Eq. (\ref{later}) shows that typically on a flow expansion time
$\beta_\theta \sim 1/\gamma^2\theta$ - 
the lateral dynamics is frozen-out for ultrarelativistic flow.

In the context of our model, the flow pattern is established in the vicinity
of  the source. 
In the subsequent motion, the flow at different latitudes drops out
of causal contact until $r\sim r_{{\rm dyn}}$,
 when connection is re-established
and lateral motion becomes possible  (the situation is analogous to the 
``Hello--Goodbye--Hello'' kinematics familiar from the theory of inflation).
As a consequence, steady state 
solutions based on the Grad-Shafranov equation are not likely to become 
valid on an expansion time-scale.
\vskip -.2 truein

\vskip -.2 truein
\subsection{Radial motion}
As a consequence of freezing of lateral dynamics
 we can separate out the radial motion, except, perhaps, 
close to the axis.
To find the simplest RFF solution during the expanding shell phase, 
we suppose that the only non-zero field components are
$B_\phi, E_\theta$. The system (\ref{FF}) can be solved
by separation of variables: 
\be
B_\phi=r^{-1}[f_1(t-r)+f_2(t+r)]g(\theta), \, \, \,
E_\theta=r^{-1}[f_1(t-r)-f_2(t+r)]g(\theta) 
\ee

We now impose two boundary conditions at the contact discontinuity,
$r=R$, specifically pressure balance and velocity matching
\be
{ B^2 \over 8 \pi \Gamma^2} = 2 \rho_ {\rm ext} c^2  \Gamma(R)^2, \, \, 
 E_{\theta} =0 \rightarrow  B(R)^2-E_\theta(R)^2=B(R)^2/\Gamma(R,\theta)^2
\ee

During the self-similar wave phase, 
the 
waves emitted by the source have all  caught up with the CD --
the magnetic bubble relaxes to a self-similar (in $t-r$ coordinates) structure.
Assuming that in the 
self-similar regime  $\Gamma^2 =\Gamma_0^2 t^{-m}$ 
(the coasting phase with the constant energy source is equivalent
to self-similar solution with power supply, $m=1$) we find
\ba &&
f_1(t-r) = \sqrt{ 16 \pi \rho_ {\rm ext} c^2 } \,
  \Gamma^2 \,  t^{ 2 m/(1+m)}\,  \left( 2 (m+1) \Gamma^2 \left(t-r \right) 
\right)^{(1-m)/
(1+m)}
\nn &&%
f_2(t+r) = \sqrt{ 16 \pi \rho_ {\rm ext} c^2 }\, (t+r)/8
\ea
The reflected (in-going wave), $f_2$, is $\Gamma^2$ times weaker than the 
outgoing -  Doppler red-shifting during the reflection from the CD.
Magnetic field:
\be
B \approx {f_1(t-r) \over r } g(\theta) =
 \sqrt{ 16 \pi \rho_ {\rm ext} c^2 } \, 
\Gamma^2 \, { t\over r}  \, \chi^{(1-m)/(1+m)} g(\theta)
\ee
 $\chi =  2 (m+1) \Gamma^2 (1-r/t)$. {\it The self-similar structure of the
magnetic  bubble looks similar to the structure of hydrodynamic
blast wave} (Blandford\&McKee 1974)!  This result is a bit surprising 
since the two systems are completely different (electro-magnetic bubble and
hydrodynamical shock wave) and solutions come from different equations
(Maxwell and Euler).
An important property of the  self-similar bubble is that 
for $1<m<3$ the magnetic field is concentrated  near CD in a thin sheath
$\sim R/\Gamma^2$.

In particular, 
for point explosion  in a homogeneous medium $\rho_{\rm ext}=$const ($m=3$)
we find
\ba &&
 B \propto {1 \over t^2 \, r \, \sin \theta\, \sqrt{ \chi}  }
, \,\,\,
 E_\Omega \propto  {   E_0  \over \sin^2 \theta} 
\ln  {  E_0  \over \rho_{\rm ex} c^2 r^3}
\nn &&
 \Gamma^2 = {1\over \sin^2 \theta} \left({ r_{nr} \over r} \right)^3
, \,\,\,
r_{nr} =\left( { 3 E_0 \over 2 \rho_{\rm ex} c^2} \right)^{1/3}
\sim 10^{18} {\rm{cm}}
\ea 
Interestingly, the 
energy  contained within the magnetic bubble decreases {\it logarithmically }
with time, so that the forward blast wave fully decouples from the
magnetic bubble only when the flow becomes trans-relativistic.
\vskip -.2 truein

\vskip -.2 truein
\subsection{ Lateral dynamics and collimation.}
At large distances, $r\sim 10^{16}$ cm, 
the flow slows down and the lateral dynamics
``un-freezes''. The flow then tries to adjust to a lateral force balance, 
which  becomes
$ g = 1 /\sin \theta$, $\Gamma \sim  1 /\sin^2 \theta$, $L_{\Omega} \sim
1/\sin^2 \theta$ when the toroidal field dominates over pressure.
Thus, at $r\sim 10^{16} cm$
 the flow relaxes to a universal lateral distribution of energy
 {\it independent} of the initial conditions.{\it The energy flow is
strongly peaked along the axis. }
\vskip -.2 truein

\vskip -.2 truein
\section{$\gamma$-ray emission}
In the RFF limit the fast speed is the speed of light and 
so no fast shocks form. If we add a limited quantity of plasma and 
use RMHD, the shocks are weak and not likely to be efficient 
particle accelerators. There are no intermediate shocks in the RFF 
limit, though rotational discontinuities can be present.

We therefore propose that the $\gamma$-ray-emitting electrons are accelerated
by current instabilities during the magnetic shell phase.  
Intense line and sheet currents are traditionally
unstable as is well-documented by laboratory experiments. There are 
two possible locations of the emission, along the poles and in the body
of the magnetic shell. We consider these in turn.
\vskip -.2 truein

\vskip -.2 truein
\subsection{Polar Emission}   
We can continue the cosmological analogy and note that as
the flow decelerates, 
 angular scales
$\sim\Gamma^{-1}$ will ``enter the horizon''. Specifically,
electromagnetic instabilities on the axis 
can grow on progressively larger scales.
The fastest growing modes are likely to be pinch, kink and helical modes.
with progressively larger scale modes becoming unstable as the flow develops,
also analogous to cosmology.

There are several ways in which these instabilities can develop. Let us 
outline one scenario that we find to be quite plausible.  If we continue to 
make the RFF approximation and ignore the inertia and pressure of the 
plasma outflow then the solution outlined above will  
have an unbalanced stress near the axis where the current flows.  This will
cause the electromagnetic field to flow towards the axis.
This inflow will be accompanied by development of current instabilities --
pinches and kinks.
  We suspect that
the nonlinear development of these instabilities will produce a turbulence 
spectrum down to a high $k$ inner scale where the particle
acceleration is able to absorb the wave energy on the wave turnover
time-scale.  Simple estimates suggest that the slope of the wave turbulence
spectrum is $3/2$ as opposed to the Kolmogorovian value of $5/3$ and that
the energy of the emitted radiation can, indeed, lie in the $\gamma$-ray 
band, (after Doppler boosting). Numerical computations of this nonlinear
development are underway (MacFadyen \& Blandford, in preparation). 
%The particle acceleration, responsible for damping the electromagnetic 
%spectrum is probably due nonlinear coupling between pairs of waves
%(\eg Blandford 1973). The actual radiation process may be synchro-Compton
%emission which can differ from synchrotron emission in its 
%polarization properties (Blandford 1972).  

One merit of this explanation is that it allows the GRB to originate at
a much larger radius, up to $\sim10^{16}$~cm than in the standard, baryonic,
intermediate shock model $\sim10^{13}$~cm.  This means that 
 the bulk  outflow
Lorentz factor can be much smaller at the emission generation radius.
 (One objection that has been raised to 
this type of model is that the it may not be possible to achieve the 
large variability on $\sim10$~ms time-scales that is observed. However,
this turns out not to be a problem if the emission zone contains 
large amplitude waves moving with speed $c$, as is likely to occur in an 
intermittent turbulence spectrum.) 
\vskip -.2 truein

\vskip -.2 truein
\subsection{Shell Emission}
The expanding shell will be preceded by a surface current emanating from the 
poles and flowing to the equator (or {\it vice versa}). It will be followed,
at a distance of the order of the shell thickness by a reverse 
current.  We strongly suspect (though have not demonstrated) that 
this electromagnetic configuration is strongly unstable and, after a few
light crossing times in the frame of the source, the shell will break up 
and create a local turbulence spectrum which can accelerate particles
as described above.  This will occur at $r\sim10^{16}$~cm. The variation,
particle acceleration and emission are much as in the polar model.
Alternatively, acceleration of external electrons 
due to inertia-induced electric fields at the contact discontinuity
(Smolsky \& Usov 1996) may lead to the production
of $\gamma$-rays.

An attractive feature 
 of the shell emission is that it predicts
$\gamma$-ray emission in all directions and few orphan afterglows. The 
strength of the burst is expected to be related to the inferred energy 
per sterad in the afterglow although large variations are indeed possible.
Burst seen at small angle (as may be inferred from achromatic
breaks in the afterglow emission) should be seen to larger distances
although a large burst to burst dispersion should be expected. It could be
that short bursts are polar and long bursts are from the shell. Simulations
of shell emission are also underway which will clarify some of these points.
\vskip -.2 truein

\vskip -.2 truein
\section{Afterglow Emission}
Under this model, the afterglow is produced during the relativistic
blast wave phase when the energy contained in the electromagnetic bubble
is transferred to the swept up circumstellar medium. However,
if the electromagnetic dynamics is as conjectured above, then this will
be reflected in the subsequent blast wave dynamics.
The form of the blast wave would reflect both the form of the driver
(magnetic bubble) and subsequent evolution of the shock.
We have  considered the
 expansion of non-spherical relativistic blast waves  in a relativistic
Kompaneets approximation (Kompaneets 1960, Shapiro 1980, Lyutikov 2002b).
We find that only extremely strongly collimated shocks, with the opening
angle $\Delta \theta \leq 1/\Gamma^2$
show  modification of profiles due to sideways expansion.
Thus, the motion of the forward shock will  be determined by the 
form of the electro-magnetic driver.
 In particular,
the energy per sterad in the shock is $ L_{\Omega} \propto 1/\theta^2$.
{\it
What is inferred as a jet is a non-spherical relativistic outflow}.
This implies that, whatever the observer 
orientation, the inferred blast wave energetics is roughly the same,
to within a logarithmic factor. The starting Lorentz factor for the 
afterglow is likely to be $\Gamma>>\theta^{-1}$. The emission should 
exhibit an achromatic break when $\Gamma$ falls to $\sim\theta^{-1}$.
Thus, all bursts are observable, although the fluency depends
strongly upon angle of observation.  This has important
 implications for the incidence of orphan afterglows and 
perhaps for the interpretation of X-ray flashes.
\vskip -.2 truein

\vskip -.2 truein
\section{Conclusion}
We have explored the  
 ``electromagnetic hypothesis'' for ultra-relativistic
outflows, namely that they are essentially electromagnetic phenomena
which are driven by energy released by spinning black holes or neutron
stars and that this electromagnetic behavior continues into the source
region even when the flows become non-relativistic.
The most striking implications of the electromagnetic
hypothesis
are that particle acceleration in the sources is due to
electromagnetic turbulence
rather than shocks and that the outflows are cold and 
electromagnetically-dominated, with very few baryons,
at least until they become strongly dissipative.

The most important prediction that is likely to be tested in the  coming years
is the form of the external shock wave generated by the 
magnetic bubble. We predict $L_{\Omega} \sim \theta^{-2}$. Observational
properties of such blast wave have been already discussed 
(Rossi at al. 2002). Current data are consistent with such 
an energy
distribution. 
Further data are expected  from optical polarization
observations, which should exhibit a  
characteristic temporary evolution as well as correlation with
intensity and the times of the jet achromatic breaks 
(Rossi at al. 2002).
\vskip -.2 truein

\vskip -.2 truein
\section*{Acknowledgements}
We are indebted to Andrew MacFadyen, Martin Rees and Elena Rossi for advice and 
encouragement. Support under NASA grant NAG 5-2837, 5-7007 is
 gratefully acknowledged.

\vskip -0.2 truein
\begin{minipage}{.45\textwidth}
%\begin{figure}[h]
\psfig{file=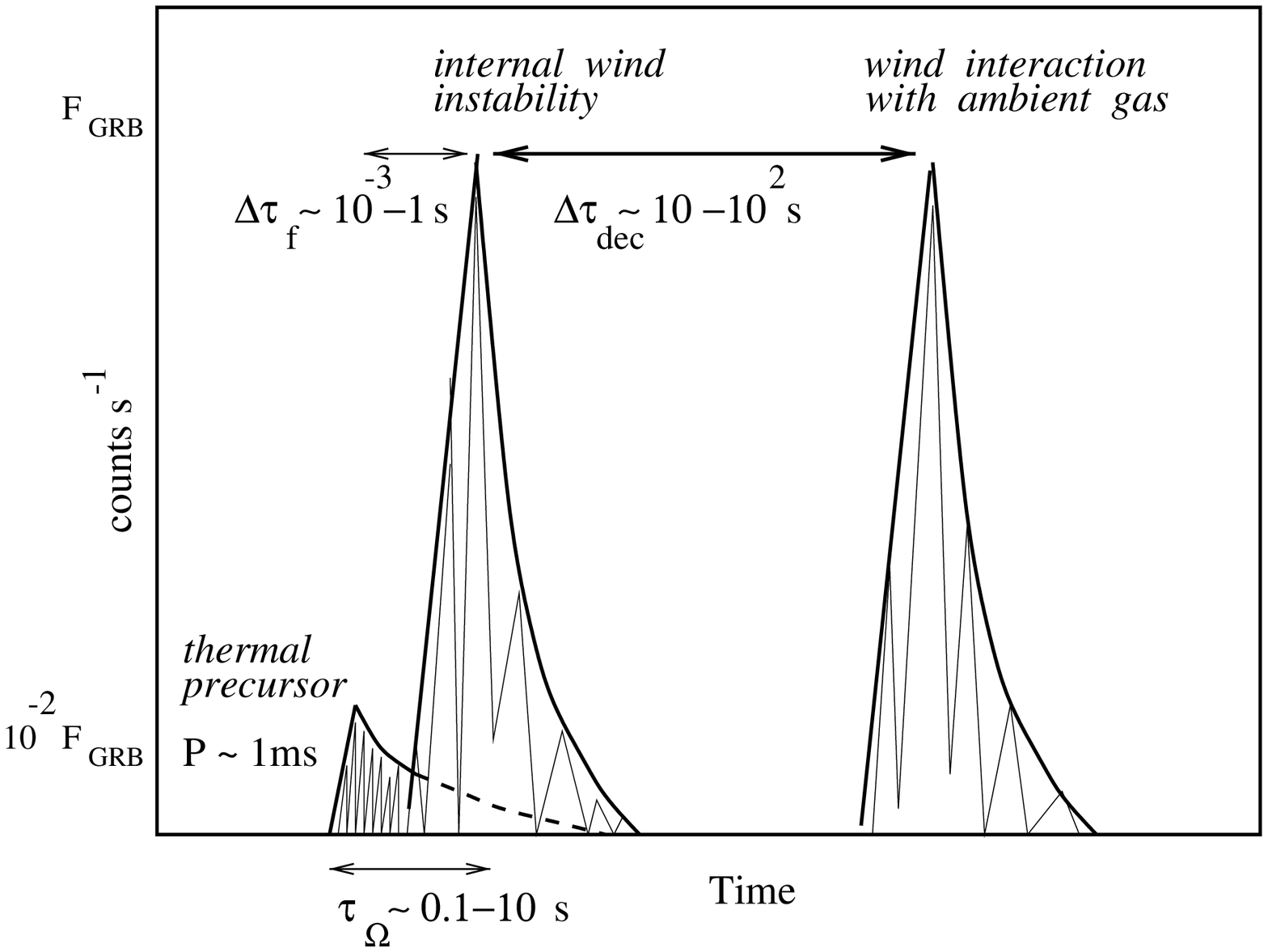,width=7cm}
%\caption
{Fig. 1. Temporal structure of GRBs (from Lyutikov \& Usov 2000)}
\label{profile}
%\end{figure}
\end{minipage}
\begin{minipage}{.45\textwidth}
%\begin{figure}[h]
\psfig{file=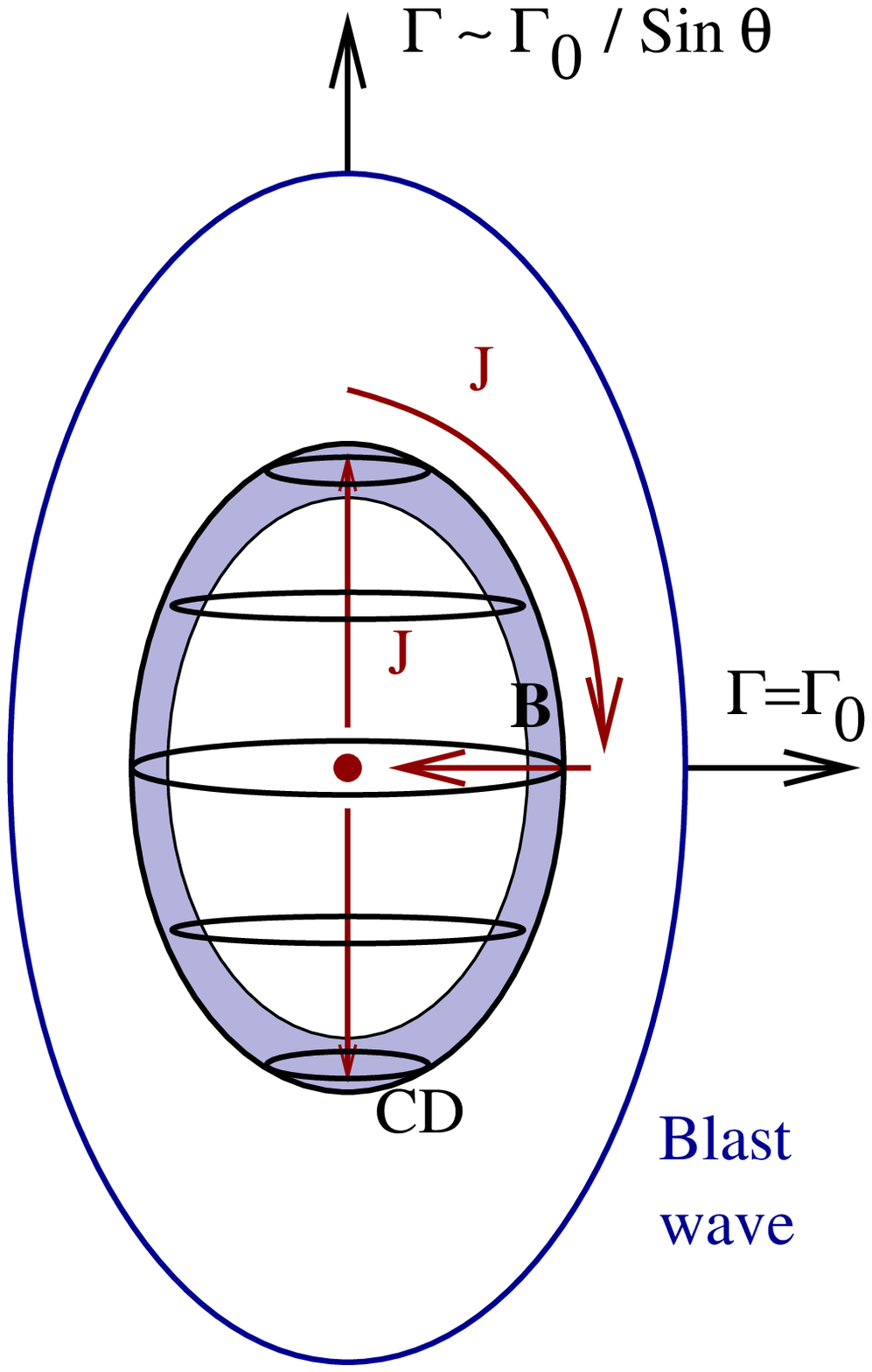,width=5cm}\\
%\caption
{Fig. 2. Magnetic fields and currents in the bubble}
\label{currentflow}
%\end{figure}
\end{minipage}
\vskip -0.4 truein

\section*{Questions}

{\bf J. Trier Frederiksen}: What are the assumptions 
of the medium that the magnetic bubble
is pushing?

{\bf M. Lyutikov}: There is  no specific requirement on the external
medium (except that it is tenuous):  it can be hot or cold,  homogeneous 
or a  power-law in radius. The internal structure  of the bubble will be 
somewhat different in these cases, but will remain qualitatively  similar.

{\bf S. Shibata}: 
What is the physics leading to "delayed reconnection" in the distance
far from the central engine?

{\bf M. Lyutikov}: Relativistic radial motion leads to an effective
freezing of lateral motion and of current instabilities,
  which becomes important only when the flow  
slows down. 

{\bf E. Berger}:
What are the relevant length scales ($10^{16}$ cm for GRB, $10^{17}$ for
afterglow, $10^{18}$ cm for non-relativistic phase) and what
determines these scales?

{\bf M. Lyutikov}:
$\gamma$-rays are emitted at $ c t   \sim \Gamma^2  c t_s \sim
 (E t_s /\rho c)^{1/4} \sim 10^{16}$  cm 
at the end of the coasting phase when  lateral dynamics of the bubble
un-freezes.
Starting from this radius, most of the  
energy is  transferred to the forward shock  which becomes
 non-relativistic  at $r \sim ( E / \rho c^2)^{1/3} 
\sim 10^{18}$ cm.


\vskip -.2 truein

\vskip -.2 truein
\begin{thebibliography}{}

%\vskip -.13  truein
%\bibitem{bl72} Blandford, R. D., 1972,  A\&A, 20, 135
%\vskip -.13  truein


%\vskip -.13  truein
%\bibitem{bl73} Blandford, R. D., 1973, A\&A, 26, 161 
%\vskip -.13  truein

\vskip -.13  truein
\bibitem{bl02} Blandford, R. D., 2002, in {\it Lighthouses of the Universe},
Gilfanov, M, Sunyaev, R, Churazov, E, eds., p. 381
\vskip -.13  truein

\vskip -.13  truein
\bibitem{blm76}Blandford, R. D. \& McKee, C. F. 1976, Phys. Fluids, 19, 1130
\vskip -.13  truein 


\vskip -.13 truein
\bibitem{fra01}Frail, D. A. et al. 2001 ApJ 534 559
\vskip -.13 truein

\vskip -.13 truein 
\bibitem{kll2002} {{Kim}, H., {Lee}, H.~K., {Lee}, C.~H.},
    2002, astro-ph/0206171
\vskip -.13 truein

\vskip -0.13 truein 
\bibitem{Kluzniak98}
Klu\'{z}niak~W., Ruderman~M. 1998, ApJ 505, L113
\vskip -.13 truein

\vskip -0.13 truein 
\bibitem{Kouveliotou} Kouveliotou C.  et al., 1998, {Nature}, 393, 235
\vskip -.13 truein

\vskip -0.13 truein
\bibitem{komp} Kompaneets A.S., 1960, Sov. Phys. Doklady, 130, 1001
\vskip -.13 truein

\vskip -0.13 truein 
\bibitem{laza02} {{Lazzati}, D., {Ramirez-Ruiz}, E., {Rees}, M.~J.},
2002, ApJ Lett., 572, 57
\vskip -.13 truein

\vskip -0.13 truein 
\bibitem{lyu02}Lyutikov, M. 2002a, Phys. Fluids,  14, 963
\vskip -.13 truein

\vskip -0.13 truein 
\bibitem{lyu02b}Lyutikov, M. 2002b, in preparation
\vskip -.13 truein

\vskip -0.13 truein 
\bibitem{lyu00} Lyutikov, M. \& Usov V.V. 2000, ApJ Lett., 543, 129
\vskip -.13 truein

\vskip -0.13 truein 
\bibitem{mac99}MacFadyen, A. \& Woosley, S. 1999, ApJ, 524
\vskip -.13 truein

\vskip -0.13 truein 
\bibitem{pana} {{Panaitescu}, A., {Kumar}, P.},  2002, ApJ, 571, 779
\vskip -.13 truein

\vskip -0.13 truein 
\bibitem{pir99} {{Piran}, T.},
    1999, Phys. Rep.,
    314,
    575
\vskip -.13 truein

\vskip -0.13 truein 
\bibitem{Pr} Preece R. 2002, private communication
\vskip -.13 truein

\vskip -0.13 truein 
\bibitem{Rossi} {Rossi}, E., {Lazzati}, D., {Rees}, M.~J., 
2002, {MNRAS}, 332, 945; also this volume
\vskip -.13 truein

\vskip -0.13 truein
\bibitem{shapiro} Shapiro, P., 1980, ApJ, 236, 958
\vskip -.13 truein

\vskip -0.13 truein
\bibitem{tommur01} {{Thompson}, C., {Murray}, N.},
2001, ApJ, 560, 339
\vskip -0.13 truein

\vskip -0.13 truein 
\bibitem{uchida} {{Uchida}, T.}, 1997, Phys. Rev. E, 56, 2181
\vskip -0.13 truein 

\vskip -0.13 truein
\bibitem{usovsm96} {{Smolsky}, M.~V., {Usov}, V.~V.},
1996, {ApJ}, 461, 858 
\vskip -0.13 truein

\vskip -0.13 truein
\bibitem{whe00} Wheeler, J. C., Yi, I., H\" oflich, P.,
Wang, L. 2000  ApJ 537 810
\vskip -0.13 truein
\end {thebibliography}
\end{document}